International Journal of Cognitive Science, Engineering, and Technology Volume 1, Issue 1, November-2013
ISSN 2347 – 8047# ENHANCING HUMAN ASPECT OF SOFTWARE ENGINEERING USING BAYESIAN CLASSIFIER

Sangita Gupta
Jain University, Bangalore.
(email- sgjain.res@gmail.com)

Suma V.
Dayanand Sagar Institutions, Bangalore.
(email sumavdsce@gmail.com)*Abstract—* **IT industries in current scenario have to struggle effectively in terms of cost, quality, service or innovation for their subsistence in the global market. Due to the swift transformation of technology, software industries owe to manage a large set of data having precious information hidden. Data mining technique enables one to effectively cope with this hidden information where it can be applied to code optimization, fault prediction and other domains which modulates the success nature of software projects. Additionally, the efficiency of the product developed further depends upon the quality of the project personnel. The position of the paper therefore is to explore potentials of project personnel in terms of their competency and skill set and its influence on quality of project. The above mentioned objective is accomplished using a Bayesian classifier in order to capture the pattern of human performance. By this means, the hidden and valuable knowledge discovered in the related databases will be summarized in the statistical structure. This mode of predictive study enables the project managers to reduce the failure ratio to a significant level and improve the performance of the project using the right choice of project personnel.**

**Keywords**- *Software Engineering, Data Mining, Project Management, Software Quality, Bayesian classification*1. Introduction

Software Engineering focuses upon generation of high quality product through well defined process. The development of software is based on an iterative process through which the code structure and functionality are improved gradually. Developing high quality software within scheduled time, cost and resources is therefore one of the major concern of any software industry. Quality can be conceived by two dimensions namely through process quality and through people quality. Since, people drive the process, the quality of software development process is controlled by the quality level of the people. Consequently, it is vital to carry the software development activities with team consisting of good skill set [8].

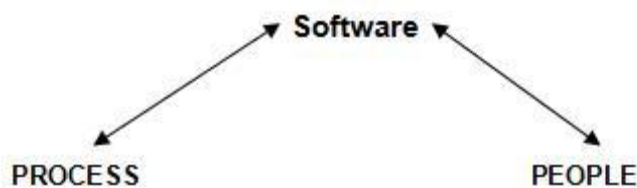

**Figure 1-Sotware Project Components**

Figure 1 infers that project quality can be achieved through cumulative efforts of people and process implementation with equal weight age[10].

In the People, Process and software project system as shown in Fig 1, the angle contributing to the Process have been well-understood, document and researched.

International Journal of Cognitive Science, Engineering and Technology | 16

However the human factor or People angle which makes significant contribution to any project needs deeper investigation. Even today, majority of project is hand-crafted (rather technology crafted) and the unpredictability of human behavior makes it difficult to bring sophistication in the trade of project engineering. The Human component , be it the motivation or energy, the emotion, the creativity and the negligence or error, largely decides success or failure of software projects or engineered products[7]. The contribution and penetration of software in the society at large is also wide and varied. As it is well known, the more the software world is developed, the more it is accepted by the software engineering community that the *people* involved in software development processes deserve more attention, not the processes themselves or technology[9].

In this spirit, the study presented by the authors of this paper attempts to highlight the world of software engineering from the perspective of the main actors that are involved in software development processes: the individual, the team, the customer, and the organization.

Thus ,software quality can be broadly defined as

Software Quality = Process Quality + people Quality
(summation over all generic framework activities)

Our investigation is to analyze the efficiency of human factor during software project development process.

Consistency in evaluations by people is always intuitive. Also, evaluation processes are generally non-parametric oriented. Intuitive decision making without any analytical method leads to wrong decisions and are unable to see hidden patterns. Therefore it is difficult to deal with such complex data without a proper parametric method and optimization technique.

This study concentrates on identifying the patterns that relate to human component in software engineering by predicting the performance of project members using parametric decision making techniques. The patterns can be generated by using some of the major data mining techniques, such as the clustering and classification technique which is used to list the employees with similar characteristics, to group the performances etc. From the association technique, patterns that are discovered can be used to associate the employee's profile for the most appropriate project, coupled with employee's attitude to performance. In classification and prediction, the pattern can be used to predict the percentage accuracy in employee's performance, behavior, and attitudes, predict the performance progress throughout the performance period, and also identify the best project members for the project.[6]

The matching of data mining problems and talent management needs is very crucial. Therefore, it is very important to determine the suitable data mining techniques. There are very few discussions on the uses of data mining related to employee's performance prediction, project assignment, employee's recruitment and many others. Due to these reasons, this study attempts to use the data mining approach for employee's performance prediction as one of the methods to enhance software quality. The purpose of this study is to determine the employees' performance by predicting their performance based on the past experience knowledge through previous performance evaluation data. In this study, the classification technique will be used for human talent prediction.

This study aims to analyze historical data and predict performance in new project. These applications can help both individual project member and manager to achieve the desired quality of project.

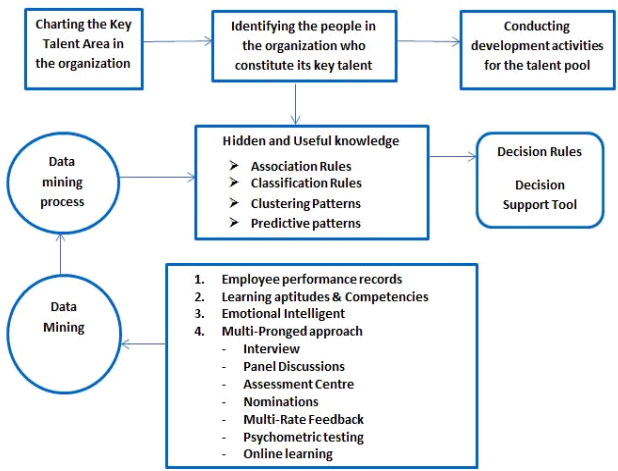

**FIG 2- OUTLINE OF RESEARCH METHODOLGY**

The above figure outlines the methodology used in this study.

The research methodology has three major phases; the first phase is the data collection process which involved the data cleaning and data pre-processing. The second phase is to generate the classification rules for the training dataset. The third phase is the evaluation and interpretation of the classification rules.

## 2. Literature survey

### A. Related work

Data mining techniques are used to operate on large volumes of data to discover hidden patterns and relationships helpful in decision making. Alternatively it has been called exploratory analysis, data driven discovery and deductive learning. Data mining access of a database differs from this traditional access in several ways: query, data and output. A data mining Algorithm is a well-defined procedure that takes data as input and produces output in the form of models or patterns. The term well-defined indicate that the procedure can be precisely encoded as a finite set of rules. The structures discovered during the data mining process can describe the entire (the most of the) set of data and they are called "models". There are also cases when the structures discovered get some local properties of the data and in that case the term of "pattern" is used[1].

The authors of [8] have used data mining for defect prediction in a software system. They have also found out through their investigation that defects occur in requirement stages rather than development stage. The authors in [5] have used non parametric methods for finding appropriate human talent in an organization.The study concentrates on decision tree algorithm Authors in [6] have done a comparative study of various data mining techniques like CART and ID3 and its efficiency for predicting the performance of new comers in a semi conductor industry.

Data mining is an emerging methodology used in project management to enhance our understanding of learning process to focus on identifying, extracting and evaluating variables of human aspects with further influence the non-functional quality attributes of functionality of projects. Data mining can be applied to a number of different applications such as data summarization, learning classification rules, finding associations, analyzing changes and detecting anomalies. Data mining prediction technique can identify the most effective factor to determine a project member's performance, and then adjusting these factors to improve the project performance. The author in [4] has used data mining technique for finding various project profitable attributes. In this paper, we attempt to discover employees' performance patterns from the existing employees' performance data using classification techniques. The techniques are chosen based on the common techniques for classification and prediction in data mining. The input variables for the process are performance factors for selected attributes (shown in fig 2); and the outcome is the employees' performance and its effect on project.The attributes for training dataset are selected based on the related factors for employee performance. These attributes are extracted from the individual factors component i.e. work outcome; knowledge and skill; individual quality; activities and contribution. The technique is explained in two steps

- Pattern recognition using Classification method
- Bayesian classification

## B. Pattern recognition using Classification method

In data mining tasks, classification and prediction is among the popular task for knowledge discovery and future plan. The classification process is known as supervised learning, where the class level or classification target is already known. There are many techniques used for classification in data mining such as Decision Tree, Bayesian, Fuzzy Logic, Support Vector Machine (SVM), Artificial Immune System (AIS), Neural Network, Rough Set Theory, Genetic Algorithm and Nearest Neighbor[3].

Classification is perhaps the most familiar and most popular data mining technique. Predication can be thought of as classifying an attribute value into one of a set of possible classes. It is often viewed as forecasting a continuous value, while classification of discrete value. Predictive modeling is the process by which a model is created or chosen to try to best predict the probability of an outcome. In many cases the model is chosen on the basis of detection theory to try to guess the probability of an outcome given a set of input data. Classification is a predictive data mining technique, makes predication about values of data using know results found from different data also called training set . Predictive models have the specific aim of allowing us to predict the unknown value of a variable of interest given known values of other variables. Predictive modeling can be thought of as mapping from an input set of vector measurements x to a scalar output y. Classification maps data into predefined groups or classes. It is often referred to as supervised learning because the classes are determined before examining the data. They often describe these classes by looking at the characteristic of data already known to belong to the classes[1]. In this study personal characteristic, educational background and work related attributes have been taken to find their effect on project personnel's performance in a software project. project personnel have been classified as good, average or poor based on the rules generated by Bayesian classifier. The next section will discuss more on the technique.

## C. Bayesian Classification

Bayesian classification has been proposed that is based on Bayes rule of conditional probability. Bayes rule is a technique to estimate the likelihood of a property given the set of data as evidence or input Bayes rule.[3] The approach is called "naïve" because it assumes the independence between the various attribute values. Naïve Bayes classification can be viewed as both a descriptive and a predictive type of algorithm. The probabilities are descriptive and are then used to predict the class membership for a target tuples. The naïve Bayes approach has several advantages: it is easy to use; unlike other classification approaches only one scan of the training data is required; easily handle mining value by simply omitting that probability. An advantage of the naive Bayes classifier is that it requires a small amount of training data to estimate the parameters (means and variances of the variables) necessary for classification. The parameters to be estimated from sample data are central moments- mean and variance. Because independent variables are assumed, only the variances of the variables for each class need to be determined and not the entire covariance matrix. In spite of their naive design and apparently over-simplified assumptions, naive Bayes classifiers have worked quite well in many complex real world situations.

### 3. Summary

The purpose of this study is to identify the important or interesting attributes and generate rules related to performance of software project personnel. The hidden knowledge from the performance evaluation is embedded into the decision support system for employees' performance prediction. Thus, the intelligent decision support system can be used to predict whether the employee is recommended for project or not.

Deploying the right talent at right project is done with the following strategy

• Identify a correct team for the project using classification technique

• Have a kick-off meeting and identify existing members

• Set goals for each project member

• Get the team trained on all major aspects related to project. Most of the activities relating to project planning are self-explanatory.

A data mining approach will raise the awareness and identify the solutions that may be available somewhere in the organization at the very start of the project. Then further decisions regarding the training for team on project and causal analysis is done. By leveraging this strategy we can achieve higher quality and productivity.

### 4. CONCLUSION

Quality assurance in software project management has raised to the top of the policy agenda in many companies. Software Data Mining is an emerging discipline, concerned with developing methods for exploring the unique software engineering data and using data mining technique and methods to better understand projects. By mining we develop new methods to discover knowledge from software project database. Lack of deep and enough knowledge in project management may prevent management to achieve quality objectives, Data mining methodology can help this knowledge gaps in project management .This study specifies the importance of Integrating workforce development with process improvement for software project success.